\documentclass[letterpaper]{article}
\usepackage{aaai22}  
\usepackage{times}  
\usepackage{helvet} 
\usepackage{courier}  
\usepackage[hyphens]{url}  
\usepackage{graphicx} 
\usepackage{booktabs}
\usepackage{natbib}  
\usepackage{caption} 
\usepackage{multicol}
\usepackage{multirow}
\usepackage{subcaption}

\usepackage{xcolor}
\newcommand{\answerYes}[1]{\textcolor{blue}{#1}} 
\newcommand{\answerNo}[1]{\textcolor{teal}{#1}} 
\newcommand{\answerNA}[1]{\textcolor{gray}{#1}}

\begin{document}

\title{NELA-PS: A Dataset of Pink Slime News Articles for the Study of Local News Ecosystems}
\author {
    Benjamin D. Horne\textsuperscript{a,b} and Maur\'{i}cio Gruppi\textsuperscript{c}
}
\affiliations {
    \textsuperscript{a}School of Information Sciences, University of Tennessee Knoxville, Knoxville, TN, USA\\
    \textsuperscript{b}Data Science and Engineering, The Bredesen Center, University of Tennessee Knoxville, Knoxville, TN, USA\\
    \textsuperscript{c}Department of Computing Sciences, Villanova University, Villanova, PA, USA\\
    bhorne6@utk.edu, mgouveag@villanova.edu\\
}

\maketitle

\begin{abstract}
Pink slime news outlets automatically produce low-quality, often partisan content that is framed as authentic local news. Given that local news is trusted by Americans and is increasingly shutting down due to financial distress, pink slime news outlets have the potential to exploit local information voids. Yet, there are gaps in understanding of pink slime production practices and tactics, particularly over time. Hence, to support future research in this area, we built a dataset of over 7.9M articles from 1093 pink slime sources over 2.5 years. This dataset is publicly-available at \url{https://doi.org/10.7910/DVN/YHWTFC}.
\end{abstract}

\section{Introduction}
Local news outlets are vital in American's information diets and to the broader media ecosystem \citep{hayes2015local, hayes2018decline, miller2018news, le2022understanding}. Yet, despite this importance, many local news outlets have closed due to financial distress \cite{rashidian2019friend}, leaving voids in local information spaces. More concretely, at least 1,800 local news outlets, roughly a quarter of all U.S. newspapers, have shut down since 2004 \citep{poynter2020deserts}. The loss in authentic local news coverage has been argued to have multiple negative impacts on American communities, such as sparse local coverage during the COVID-19 pandemic \citep{poynter2021covid, joseph2022local} and a lack of local watchdogs, which may make local governments less efficient and transparent \citep{miller2018news}.

A concern within this bigger picture is that inadequate information sources, or even malicious information sources, will fill these growing information voids. One particular type of information source that may fill local information voids are pink slime news outlets. Pink slime news outlets produce low-quality, often automated, content that is framed as local news \cite{bengani2019hundreds}. Of the few works that have studied pink slime news, they have shown that these outlets are often members of larger, coordinated networks that are controlled by central entities \cite{bengani2019hundreds, bengani2020election}, and that they ``prioritize the publication of state and national partisan content at the expense of local news'' \cite{royal2022local}. Furthermore, these outlets have been said to lack funding transparency and may ``engage in pay-for-play political influence operations'' \cite{bengani2019hundreds, alba2020local, royal2022local}. This coordinated promotion of partisan agendas framed as authentic local news is particularly worrisome as Americans tend to trust local news \cite{nyhan2019americans, gottfried2021partisan}.

Despite the potential damage to American's information diets being done by pink slime outlets, gaps exist in our understanding of these outlets operations and content production, particularly over long periods of time. Hence, to better support the study of pink slime news tactics, we construct a near complete dataset of over \textbf{7.9M articles} from \textbf{1093 pink slime news outlets} over \textbf{2.5 years}. In this paper, we describe the dataset's collection methods, publicly-available formats, and provide a brief, descriptive comparison of pink slime news and authentic local news.

\paragraph{Broader Perspectives, Ethics, and Competing Interests}
We hope that work done with this dataset will create knowledge and tools to better support local news in the United States. While this dataset makes up only one part of this bigger picture, it covers a currently unfilled resource gap in this research agenda. Therefore, we believe that making this dataset available for research use can be a public benefit. NELA-PS only contains freely and publicly available articles. None of this data was scraped from outlets that have paywalls or other pay-to-access services. We also considered removing the author bylines from the dataset, but as shown by \citet{royal2022local}, the author bylines are critical in understanding the use of automation, syndication, and outsourcing done by these networks. Therefore, for the sake of provenance and research use of the data, we kept the author bylines as they are on the public webpages. Other considerations can be found in the Paper Checklist. The authors declare no competing interests.

\section{Related Work and Data Use Cases}
While relatively little work has focused on pink slime news, the phenomenon is not unlike large telecommunications conglomerates taking ownership of many U.S. local news outlets. For example, critics have pointed out that when local news is apart of a national conglomerate, those outlets tend to cover the national news over local news \cite{martin2019local}. These outlets may also converge on style and sourcing. For example, \citet{hedding2019sinclair} demonstrated that local television news owned by Sinclair Broadcast Group produced more stories with partisan sources and ``dramatic elements''. Similarly, \citet{martin2019local} showed that ownership by a conglomerate shifted coverage of local news to be more ideologically right leaning. More broadly, the nationalization of local news is a major concern of media scholars as it has been associated with political polarization \cite{melusky2020local, darr2021home}, and it changes both the amount and selection of local news covered \cite{toff2021social}.

From what we understand so far, pink slime networks are akin to this phenomenon as they produce the much of the same, often national, content across locations, and they (at least those networks that have been studied so far) tend to produce more ideologically right leaning content and topics. For example, according to analysis done by \citet{royal2022local} on the Metric Media pink slime network, 76\% of all human written content during the 2.5 months studied (between November 16, 2020 to February 1, 2021) was on U.S. election fraud. After the 2020 general election, stories on absentee ballot rejection rates were published across the network. The key difference between telecommunications conglomerates and pink slime networks is the heavy use of automation and outsourced authors. Again, according to analysis done by \citet{royal2022local} on the Metric Media pink slime network, the vast majority of content was automated and the median human author published stories across 9 states. These automated stories were largely not substantial local news \cite{royal2022local}. As discussed by \citet{bengani2019hundreds}, many of the automated services on these sites ``relied on data releases from federal programs'' rather than producing novel local information.

Pink Slime networks are different than partisan national news organizations, such as Brietbart or NewsMax, in that they mascaraed as neutral, local organizations and are networked. Although, the political and financial backing of these organizations may be equally opaque \cite{bengani2020election}. 

NELA-PS is meant to support the continued research of these pink slime networks, as we still do not understand the long-term strategies of these outlets. For example, while \citet{royal2022local} provides a clear case study of 2020 U.S. election fraud coverage by these sources, questions remain about what other events cause an influx of both human written and automated partisan coverage among the generic automated content normally published by these outlets. NELA-PS not only allows for similar in-depth case study analysis but also longitudinal comparisons to other types of media, such as independent local news, conglomerate local news, national mainstream media, or hyper-partisan blogs. We provide a list of other datasets and resources that can be paired with NELA-PS for this type of analysis below. 

\begin{figure}[ht!]
    \centering
    \includegraphics[width=5.75cm]{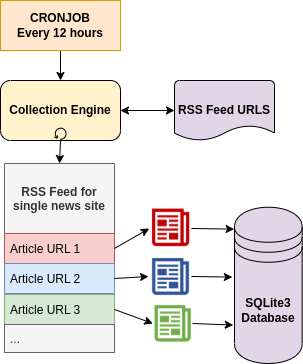}
    \caption{Flow diagram of article collection system. Note the collection system is the same system that was developed in \citet{norregaard2019nela} and used in \citet{horne2022nela}. This flow graphic is from \citet{horne2022nela}.}
    \label{fig:flow}
\end{figure}

\begin{table*}[ht!]
    \centering
    \begin{tabular}{c|c|c|c|c}
    \toprule
        \textbf{Network} & \textbf{\# of Outlets} & \textbf{\# of Articles} & \textbf{\# of Locations (States)} & \textbf{\# of IP Addresses}\\\midrule
        Metric Media & 967 & 6,996,161 & 50 & 5\\
        Metro Business Network (Franklin Archer) & 51 & 415,983 & 51 & 1\\
        LGIS & 35 & 415,296 & 3 & 1\\
        Record & 12 & 56,182 & 10 & 1\\
        Franklin Archer & 3 & 19,390 & 3 & 3\\
        American Catholic Tribune Media Network & 6 & 6,395 & 6 & 1\\
        Locality Labs & 3 & 1,295 & 1 & 2\\
        Local News Network (Franklin Archer) & 15 & 247 & 7 & 5\\
        Organisation & 1 & 16 & 1 & 1\\\midrule
        \textbf{Total} & \textbf{1,093} & \textbf{7,910,965} & \textbf{54} (unique) & \textbf{16} (unique) \\\bottomrule
    \end{tabular}
    \caption{The number of outlets per pink slime network. Note, Organisation means the top level organization website, which often do not produce any news. We chose to scrape there RSS feeds in case this changed during the time frame. There are 54 unique locations as one is national (U.S.) and two others are unknown. The rest are the 50 U.S. states and the District of Columbia. IP Addresses were gathered after the news article data collection in 2024, hence they may have changed over collection time frame. For more information about the outlet to network classification, please see \citet{bengani2020election}.}
    \label{tab:outlets}
\end{table*}

\begin{figure*}[ht!]
    \centering
    \begin{subfigure}{.33\textwidth}
        \centering
        \includegraphics[width=\textwidth]{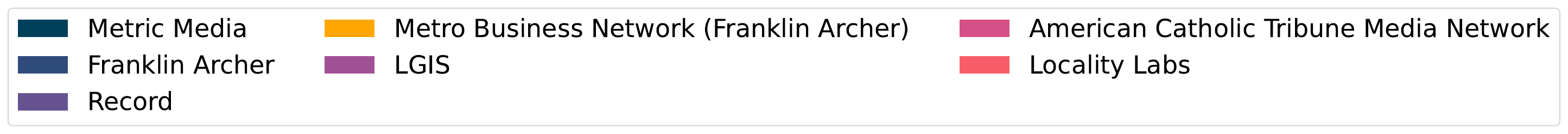}\\
        \includegraphics[width=\textwidth]{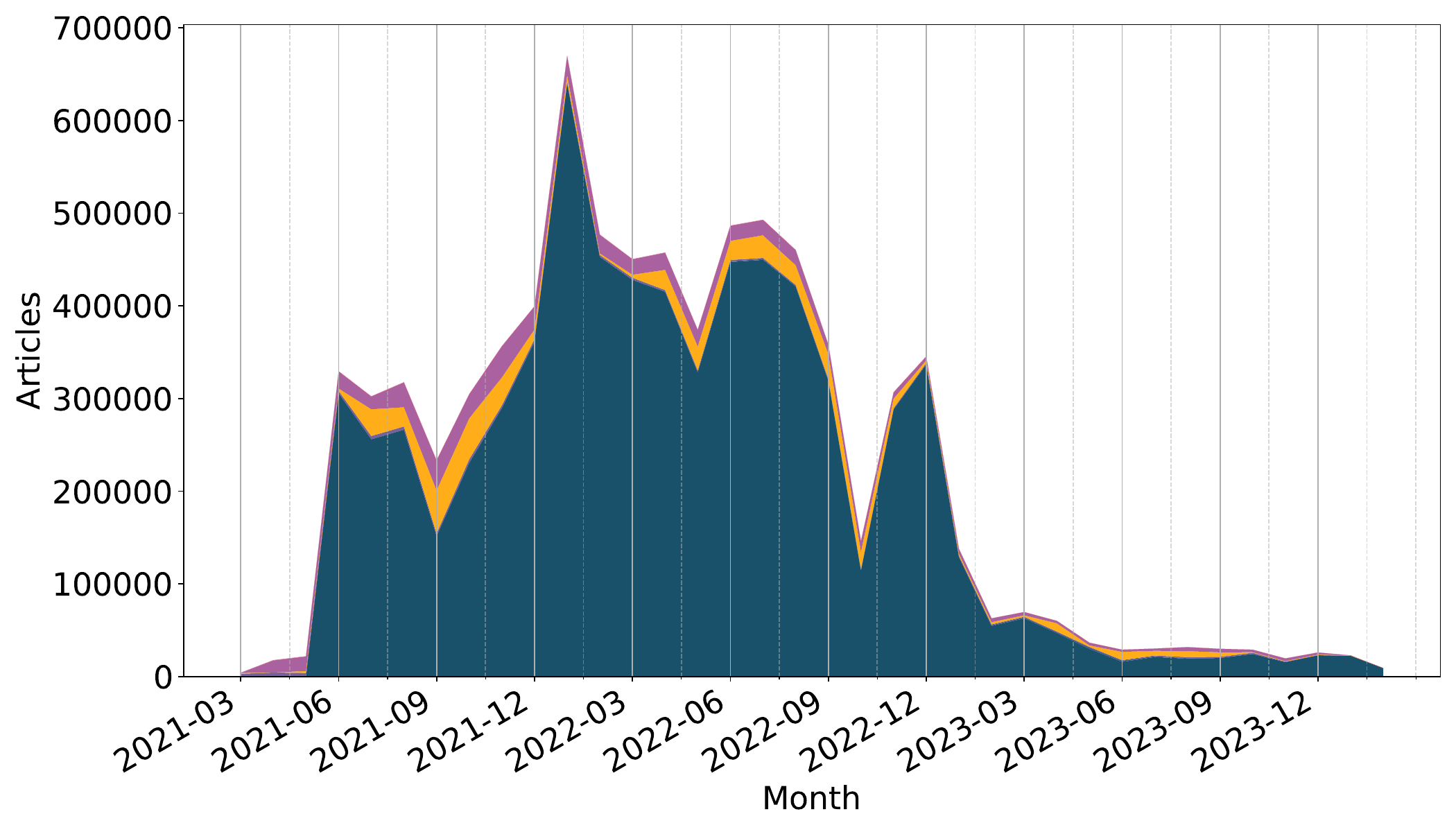}
        \caption{Articles Per Month by Each Network}
        \label{fig:time}
    \end{subfigure}
    \begin{subfigure}{.33\textwidth}
        \centering
        \includegraphics[width=\textwidth]{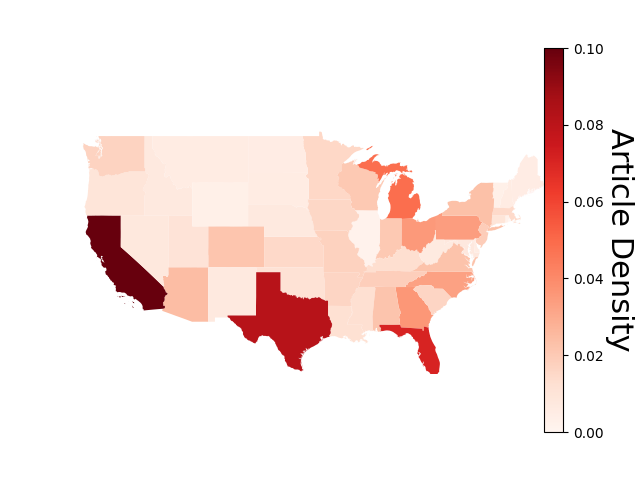}
        \caption{Article Density}
        \label{fig:map1}
    \end{subfigure}
    \begin{subfigure}{.33\textwidth}
        \centering
        \includegraphics[width=\textwidth]{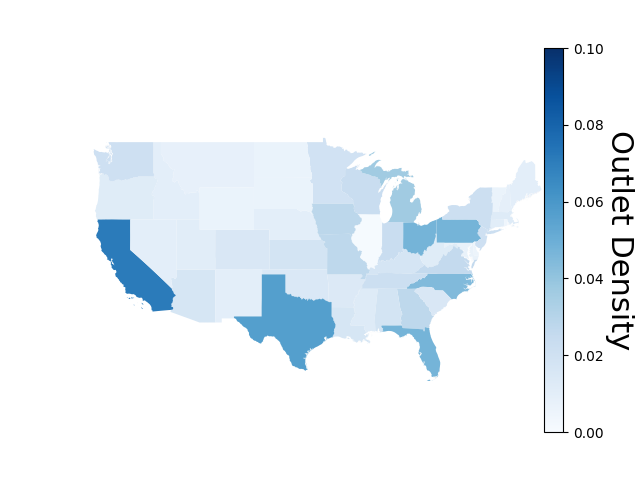}
        \caption{Outlet Density}
        \label{fig:map2}
    \end{subfigure}
\caption{(a) Number of articles published per month per pink slime network. Note, some networks produce so little in comparison to Metric Media, they are not easily seen in the figure. (b) The article density per state over the full dataset, where darker red is higher density. (c) Outlet density per state, where darker blue is higher density. Not shown in both maps are Alaska with 8 outlets and 0.47\% of the articles and Hawaii with 6 outlets and 0.24\% of the articles.}
\end{figure*}

\section{Related Datasets and Resources}
There are several related news datasets that can be used in conjunction with NELA-PS to answer various research questions. First, the \textbf{NELA-Local} dataset can serve as comparison to work done using NELA-PS \cite{horne2022nela}. NELA-Local is a dataset articles from 313 authentic U.S. local news outlets between April 2020 and December 2021. The dataset contains a range of county-level metadata, which could be aggregated to the state level for use with the NELA-PS dataset. Similarly, the \textbf{NELA-GT} datasets, which are yearly news datasets that cover national and fringe news \cite{norregaard2019nela,gruppi2020nela,gruppi2021nela}, can be used in conjunction with NELA-PS, depending on the research question. The hope is that given that both of these datasets are collected and stored using the same methods as NELA-PS, mixing and matching data across these sets should be easy.

Second, \textbf{Media Cloud} is a platform that has provided news data from a wide range of national and international outlets since 2011 \cite{roberts2021media}. Similar to the NELA datasets, Media Cloud collects news article data and publication metadata, but it does not provide the full text data. The metadata collected by Media Cloud can be a valuable resource to draw comparisons between pink slime and other types of news media.

Another valuable set of resources and data that can be used with NELA-PS are from \textbf{US News Deserts Database} \cite{abernathy2016rise} at UNC and the \textbf{Local News Initiative}\footnote{\url{localnewsinitiative.northwestern.edu}} at Northwestern. These projects have tracked closures of U.S. local news outlets and provide historical local news data by request. Both the data and analysis done in these projects can support research done with the NELA-PS dataset.

Lastly, while the data may not be publicly-available, it is worth noting that the data in NELA-PS should be approximately a continuation of the 2.5 months of data used in work by \citet{royal2022local}, as a similar collection method is used across overlapping pink slime networks prior to the start of NELA-PS. Specifically, the data used in \citet{royal2022local}'s work covers 999 pink slime outlets from November 16th, 2020 to February 1, 2021. The NELA-PS collection start soon after this time frame on March 1st, 2021 and covers 1093 outlets, many of which should overlap with the outlets used by \citet{royal2022local}. Both datasets contain data from the Metric Media and LGIS networks.

\section{Collection Methods}
\subsection{Pink Slime Outlets}
First, we gathered a list of pink slime networks from the invaluable work done by \citet{bengani2019hundreds,bengani2020election}. These networks included Metric Media, The Record Network, LGIS, Franklin Archer, Metro Business, Locality Labs, and the American Catholic Tribune Media Network. As described by \citet{bengani2019hundreds,bengani2020election}, the lines between these networks are blurry at best. Nonetheless, we gathered URLs for each outlet listed on these networks webpages and documented which network they come from. In addition, we document the location per outlet as determined by \citet{bengani2020election}.

\subsection{News Article Collection}
We searched for the RSS feeds in each of the outlets' websites, which contained all of the articles by each source. The RSS feeds were searched by sending an HTTP request to each outlet domain appended by one of the following common RSS feed addresses: `/rss', `/feed', `/rss.xml', `/atom', `/?feed=rss2', `/stories.rss'. If a valid response was obtained and the content could be parsed into XML, the respective address was included in the list of feeds to scrape.

Once all websites were processed, we set up the data collection to run twice a day, going over each of the collected feeds and scraping all articles in it that had not been collected before. This process follows the same approach as other NELA datasets \cite{norregaard2019nela, gruppi2020nela, horne2022nela}.

More specifically, as shown in Figure \ref{fig:flow}, the collection engine takes a list of RSS feeds as input, opens each news article URL in each RSS feed, scrapes the full article text and metadata from each webpage, and stores this data in an SQLite3 database. This process is ran every 12 hours to ensure all published data is captured. This collection of pink slime articles ran every day between March 1, 2021 and January 4, 2024, resulting in 7,910,965 articles from 1093 sources. While our original list of sources contained 1203 outlets, 110 of them did not have active RSS feeds, 6 of which were the top-level organization pages. Furthermore, some of the networks and outlets stopped producing articles during the collection. Given the stability of the live RSS feed collection method and our collection machines, we are confident that this dataset contains nearly every article produced by these sources over the 33 months.

\subsection{IP Address Sharing Data}
Following the findings of \citet{bengani2019hundreds}, we collected the IP addresses from all the websites in the list. The IP collection took place on January 9th, 2024. The results of this IP collection align with the prior work by \citet{bengani2019hundreds}, showing a massive overlap in the IP addresses used by the websites in NELA-PS. Across the full dataset, there were only 16 unique IP addresses for the 1072 outlets that still existed on January 9th, 2024. We include this data in NELA-PS as an anonymized identifier representing the IP address (i.e. IP\_Address\_10).

\section{Data Description}

To give readers a better sense of the dataset and how it aligns with the prior work by \citet{bengani2019hundreds,bengani2020election}, we describe the data in several ways. First, in Table \ref{tab:outlets}, we show the number of outlets and number of articles per pink slime network. Note, these networks may overlap. For example, from \citet{bengani2019hundreds}'s work, there is evidence that all of these networks are connected, although the extent to which each are connected is unclear. The vast majority of these outlets fall under the Metric Media network.

Second, in Figure \ref{fig:time}, we show the number of articles per month in NELA-PS, broken down by the network those articles fall under. As expected, Metric Media dominates the production throughout the timeline.  Notably, this production timeline is inconsistent throughout. This inconsistency is not due to missing data, but rather production changes by these outlets. While prior work has not examined production timelines of this size, it has been suggested that targeted events may drive the production patterns of these networks \cite{royal2022local}. Future work can investigate the drivers of these production changes.

Lastly, in Figures \ref{fig:map1} and \ref{fig:map2}, we show the density of articles and outlets per state. These maps closely follow the state statistics produced by \citet{bengani2020election} on a smaller dataset. California had the most outlets, with 78 of the total outlets, followed by Texas with 62 and Pennsylvania, Ohio, and Florida with 52 outlets each. The number of articles produced per state is highly correlated with the outlet density per state (0.923 correlation coefficient).

\begin{figure}[ht!]
    \centering
    \subfloat[\centering LN - Articles Per Source]{{\includegraphics[trim=0 0cm 0cm 0,clip,width=4.2cm]{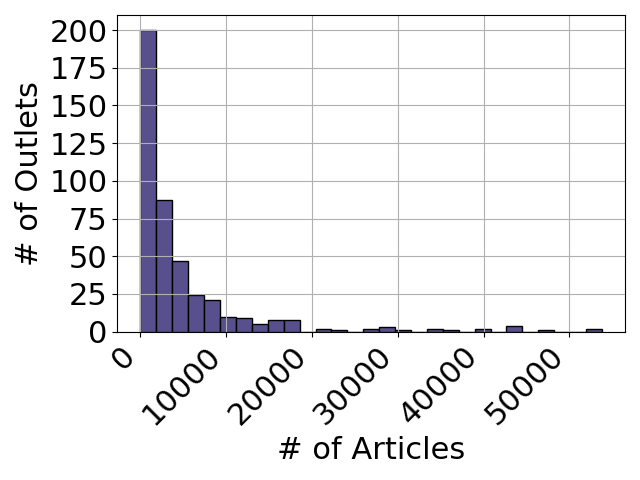}} }
    \subfloat[\centering PS - Articles Per Source]{{\includegraphics[trim=0 0cm 0cm 0,clip,width=4.2cm]{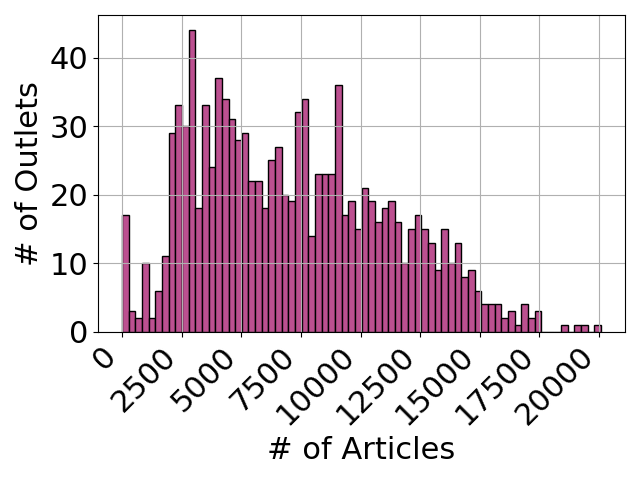} }}\\
    \subfloat[\centering LN - Articles Per Day]{{\includegraphics[trim=0 0cm 0cm 0,clip,width=4.2cm]{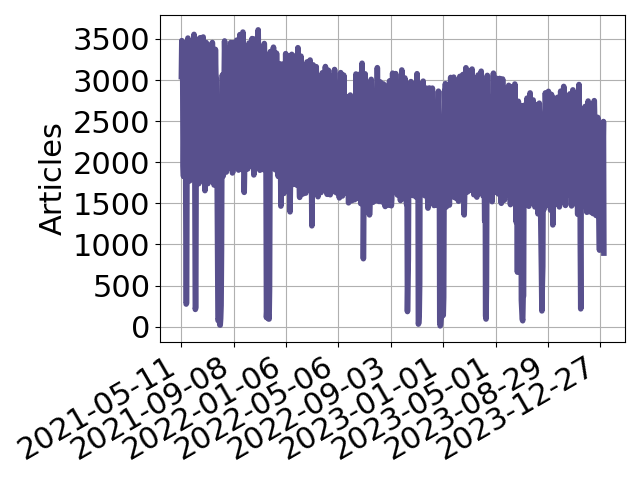}} }
    \subfloat[\centering PS - Articles Per Day]{{\includegraphics[trim=0 0cm 0cm 0,clip,width=4.2cm]{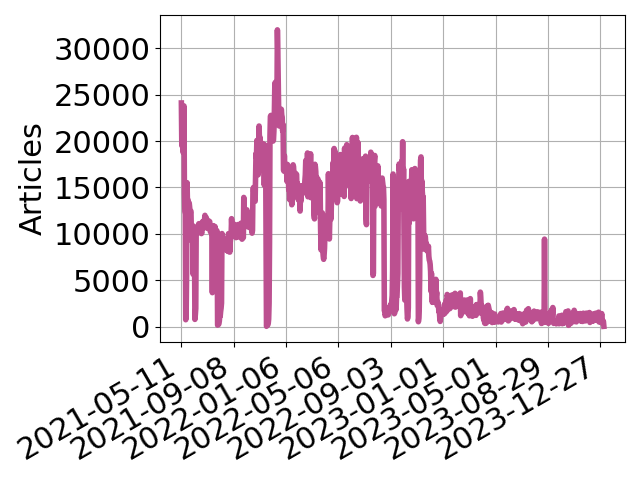} }}\\
    \subfloat[\centering LN - Words Per Article]{{\includegraphics[trim=0 0cm 0cm 0,clip,width=4.2cm]{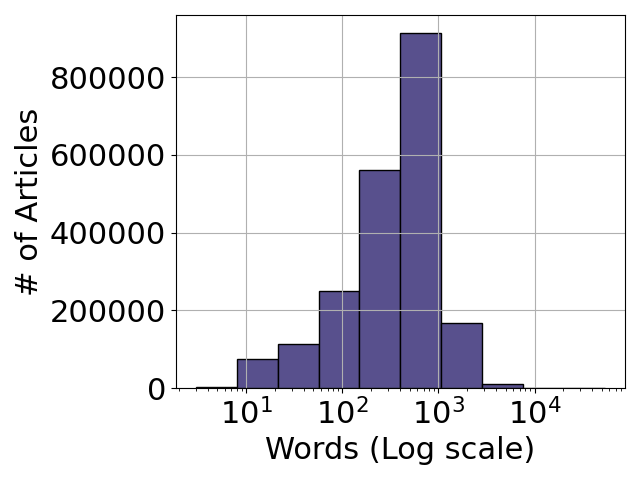}} }
    \subfloat[\centering PS - Words Per Article]{{\includegraphics[trim=0 0cm 0cm 0,clip,width=4.2cm]{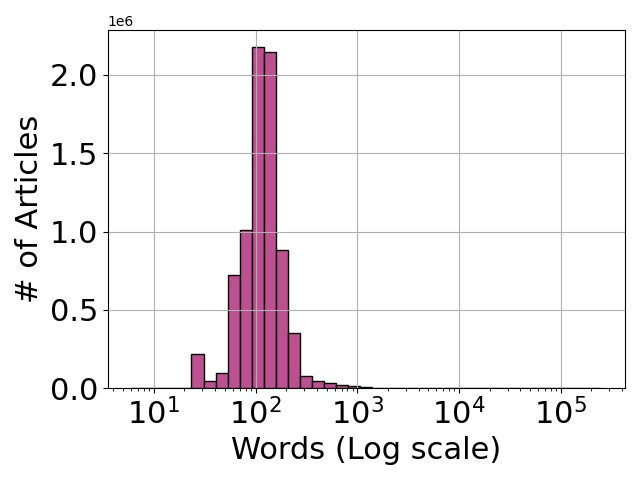} }}\\
    \caption{Comparisons between the NELA-Local (LN) dataset (column 1) and the NELA-PS (PS) dataset (column 2) over the same time frame. In (a) and (b), we show the distributions of the number of articles per outlet. In (c) and (d), we show the timeline of the number of articles published per day. In (e) and (f), we show the log distribution of the number of words per article. Note the difference in the x and y axes scales between the columns.}%
    \label{fig:dist}%
\end{figure}

\begin{figure*}[ht!]
    \centering
     \subfloat[\centering 05/21 to 09/21]{{\includegraphics[width=6cm]{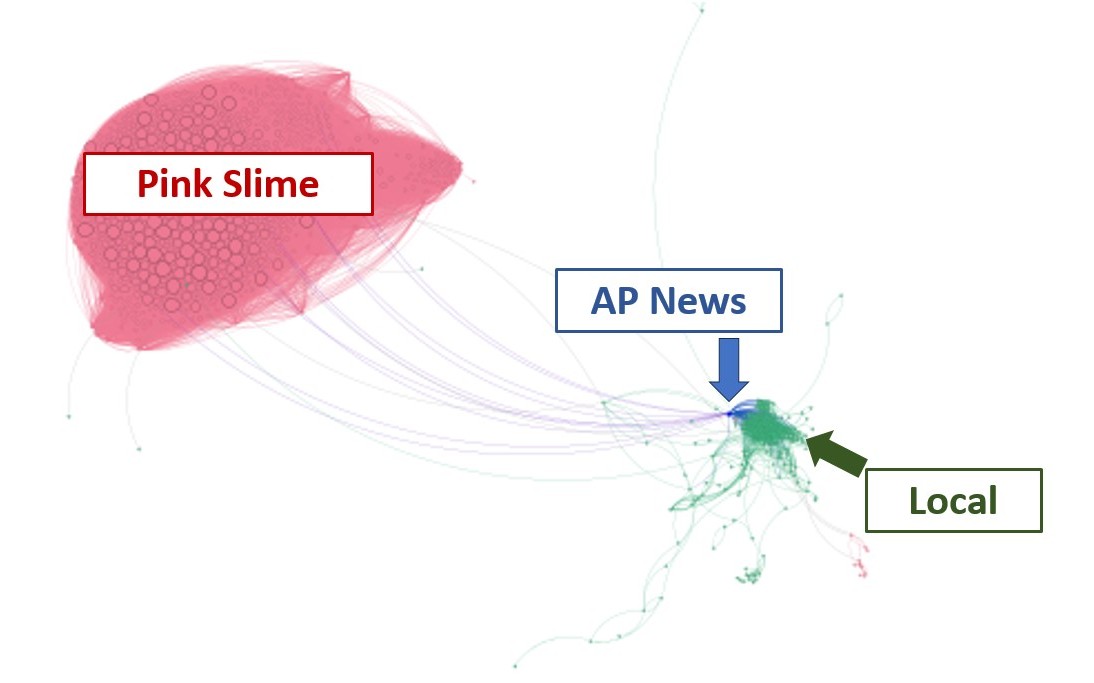} }}%
    \subfloat[\centering 10/21 to 02/22]{{\includegraphics[width=6cm]{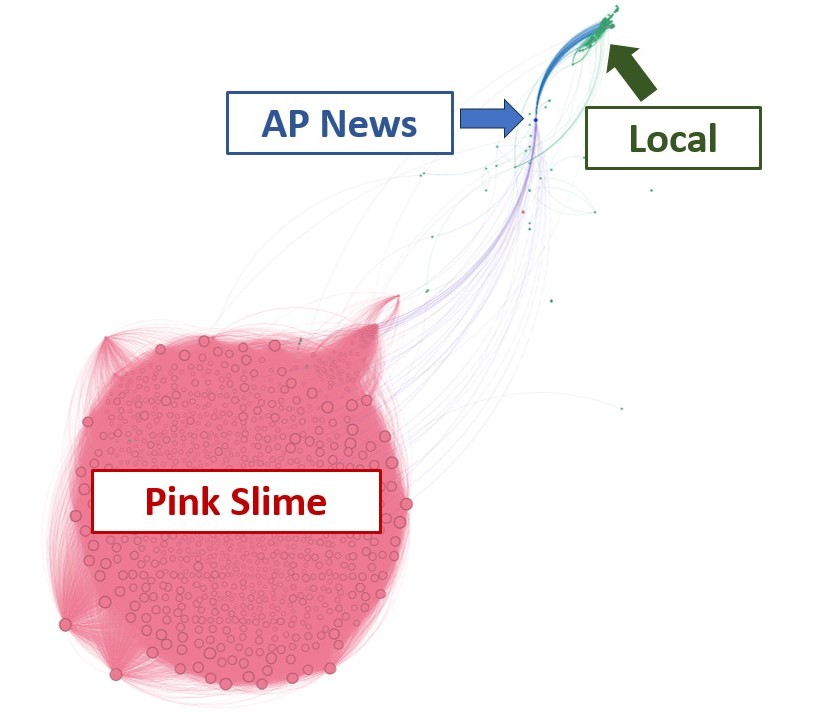} }}%
    \subfloat[\centering 03/22 to 08/22]{{\includegraphics[width=6cm]{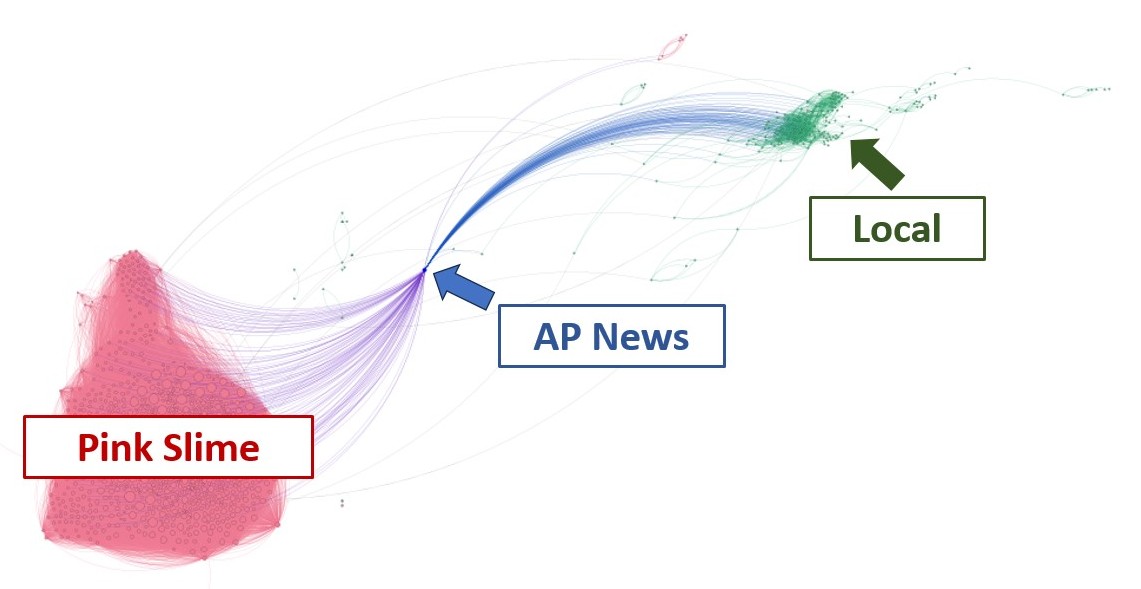} }}\\
    \caption{Content Sharing Network across time, where colors represent if the outlet is a pink slime outlet (colored and annotated in red) or authentic local news outlet (colored and annotated in green). Nodes represent outlets. Node size is based on how many articles are copied from that source and edges are directed weighted edges in the direction of information flow (A $\rightarrow$ B, means B copied from A). We show these networks across three 5-month subsets: articles published between May 2021 and September 2021, between October 2021 and February 2022, and between March 2022 and August 2022. Note, the primary bridge node between the pink slime and local news groups is AP News (colored and annotated in blue). The layout of each network is generated by the Force-atlas-2 layout algorithm in Gephi \cite{bastian2009gephi, jacomy2014forceatlas2}. This figure is best viewed in color.}%
    \label{fig:csns}%
\end{figure*}

\subsection{A Brief Comparison Between Pink Slime and Authentic Local News}
To further demonstrate that the data collected in NELA-PS is indeed different than authentic local news, we briefly draw comparison to an extended version of the NELA-Local dataset \cite{horne2022nela}. NELA-Local is a dataset articles from 313 U.S. local news outlets. The original dataset covered local news between April 2020 and December 2021. We extend and cut this dataset to create an overlapping subset with NELA-PS (May 2021 to December 2023). This extend version of NELA-Local contains 2,193,331 articles. Importantly, the outlets in NELA-Local were vetted to be authentic local news outlets \cite{horne2022nela} and, as expected, zero outlets overlap between the two datasets.

We compared these datasets in a few ways. First, we compared three distributions of production across the datasets: articles per source, articles per day, and words per article (Figure \ref{fig:dist}). This juxtaposition demonstrates the automated behavior found across NELA-PS. Most notably, shown in Figure \ref{fig:dist}(a) and (b), the distribution of the number of articles per source for NELA-PS is nearly a uniform distribution, while the distribution for NELA-Local is the skewed distribution that we expect from human behavior \cite{muchnik2013origins}. In addition, we see that the number of articles produced per day in NELA-PS is magnitudes greater than the articles produced per day in NELA-Local. These timelines also show the relatively consistent publication pattern of authentic local news versus a spiking publication pattern of pink slime news (aligning with Figure \ref{fig:time}). For example, between May 2021 and May 2023, pink slime production ranged widely, from 30,000 to 1000 articles per day. However, for almost all of 2023, production ranged between approximately 1000 articles a day to  5000 articles per day. On the other hand, local news consistently produces between 1500 and 3500 articles per day, with some exceptions. Lastly, on average, the number of words per article is more in NELA-Local than in NELA-PS (517.00 words versus 133.78 words on average, respectively), reflecting the short, automated articles described by \cite{royal2022local}.

Second, we created Content Sharing Networks (CSN) across three time subsets of NELA-PS and the extended version of NELA-Local. These CSNs were built using the method described in \cite{horne2019different}. In short, we built a TFIDF matrix for every five days with in the three five-month subsets, created directed edges between articles which had a cosine similarity greater than $0.85$ (ordered by publication times), and aggregated these article-level links into outlets-level links. This process created three directed networks in which edges are weighted, directed edges representing the probability of articles being copied from one outlet to another, and nodes represent news outlets. The visualization of these networks shown in Figure \ref{fig:csns}. More details on this method can be found in \citet{horne2019different}.

Pink slime outlets shared much of their content with other pink slime outlets, as shown by the large, tightly connected red communities in Figure \ref{fig:csns}. While authentic local news shared very little of their content with other local news outlets. Bridging the pink slime and local news communities is AP News, a commonly syndicated national news outlet \cite{horne2019different}. Although small in comparison to the number of copies within the pink slime community, there is both national and local news being imported into the pink slime community. Namely, across the three networks, 12,447 articles from AP News were copied by pink slime outlets and 36,063 articles from authentic local news outlets were copied by pink slime outlets. Perhaps unsurprisingly given the work by \cite{bengani2019hundreds,bengani2020election}, despite NELA-PS covering multiple different pink slime networks, the community structure does not reflect this. This lack of community structure supports the blurry ownership and operation behind these networks. These patterns were consistent across all three time slices of the datasets.

\section{Data Formats and Distribution}
To ensure ease of access to many different scientist, we provide three different formats for the data: SQLite3 database, CSV, and JSON. 

\subsection{SQLite3 Database}
The database schema follows a simple, single table format with the following columns:
\begin{itemize}
    \item id - Unique identifier for each article. This identifier is formatted the same as other NELA datasets: the source name, date, and first 100 characters of the title separated by two dashes.
    \item date - The date of article publication according to the article webpage. This date is formatted as YYYY-MM-DD.
    \item source - The source of the article, normalized to be all lower case with no spaces.
    \item network - The network in which the source belongs to (i.e. Metric Media, etc.). This network mapping comes from \citet{bengani2020election}.
    \item ip2024 - An source-level identifier to capture what outlets used the same IP addresses in 2024.
    \item location - The state in which the outlet is ``located'' in. This state mapping comes from \citet{bengani2020election}.
    \item title - The title of the article.
    \item content - The full textual content of the article.
    \item author - The author byline, if available on the webpage.
    \item url - The original URL of the article.
    \item published - Publication date time string as provided by source.
    \item published\_utc - Publication time as unix timestamp.
    \item collection\_utc -  Collection time as unix timestamp.
\end{itemize}

We provide example code for data extraction and use in the following GitHub repository: \url{https://github.com/MELALab/nela-pink-slime}. 

\subsection{CSV}
We replicate the yearly SQLite3 format as CSV files. The Comma Separated Value (CSV) format follows the same structure as the SQLite3 columns. We also provide sample data file with a random 200 articles as a CSV file.

\subsection{JSON}
In addition to the SQLite3 and CSV formats, the data is also provided in JavaScript Object Notation (JSON) form. In this version, the data is split into several JSON files, each corresponding to a news outlet, containing all articles published by that source.

\subsection{Distribution and Maintenance}
The dataset is publicly-available in the NELA Harvard Dataverse repository under CC BY-NC 4.0 License\footnote{\url{https://dataverse.harvard.edu/dataset.xhtml?persistentId=doi:10.7910/DVN/YHWTFC}}. While the dataset described in this paper will remain static for provenance, the dataset will be supported by the authors. The authors have a track record of maintaining and supporting datasets for the research community \cite{norregaard2019nela, gruppi2020nela, horne2022nela, trujillo2022mela, horne2022psycho}. Any updates to the dataset will be documented and deposited in a separate repository.

\subsection{FAIR Principles}
NELA-PS follows FAIR principles. First, the data is \textit{Findable}, as it is persistently stored on Harvard Dataverse and all metadata clearly include the identifier of the data they describe. The SQLite3 format is searchable using SQL or with a SQL browser (we recommend DB Browser for SQLite\footnote{https://sqlitebrowser.org/}). The data is \textit{Accessible} and \textit{Interoperable} as we provide three widely-used, standard formats (SQLite3, CSV, JSON) with example code for data extraction. The data is retrievable through Harvard Dataverse’s GUI. Furthermore, we provide a list of other datasets that can be integrated with NELA-PS to answer a variety of research questions. The data is \textit{Re-usable} as it is released with accessible data usage license and the data contains both URLs and documentation to maintain provenance.

\section{Data Use Guidance}
The data in NELA-PS is meant to be used for responsible research on media ecosystems. There are several important considerations regarding how this data is used. First, this data should not be used to train ''fake news'' classifiers or be used with the assumption of ground truth. While there is evidence of malicious intent behind these pink slime networks, this does not mean the news reported by these outlets should be considered ``fake'' or ``false''. Much of the information produced by these outlets is a mix of authentic news syndicated from national and local news outlets, along with automated articles that scrape from various real data sources. While there is certainly a large amount of partisan framing in the data, and there may even be false information in the data, the ground truth behind much of the dataset is likely nuanced. Second, since there is potential for offensive and false content in this dataset, introducing risks when using this dataset to train Large Language Models for conversational agents, hence, we argue that this data should not be used for that purpose.  

\section{Conclusion}
In this paper, we describe the NELA-PS dataset, a publicly-available dataset of pink slime news articles and metadata. The dataset covers 7.9M articles from 1093 outlets over 33 months between March 2021 and January 2024. It includes metadata about author bylines, outlet's audience location, and outlet IP addresses. The goal of the dataset is to support research on the long-term production strategies of these pink slime networks and to better understand their place in local information spaces. The dataset can be found at \url{https://doi.org/10.7910/DVN/YHWTFC}.

\bibliography{scibib}

\section{Paper Checklist}

\begin{enumerate}

\item For most authors...
\begin{enumerate}
    \item  Would answering this research question advance science without violating social contracts, such as violating privacy norms, perpetuating unfair profiling, exacerbating the socio-economic divide, or implying disrespect to societies or cultures?
    \answerYes{Yes, the collection and sharing of this dataset fills a resource gap in the study of media ecosystems without violating social contracts. NELA-PS only contains freely and publicly available articles. None of this data was scraped from outlets that have paywalls or other pay-to-access services.}
  \item Do your main claims in the abstract and introduction accurately reflect the paper's contributions and scope?
    \answerYes{Yes, the abstract accurately represents the dataset presented in the paper and makes no claims outside of the scope of this paper.}
   \item Do you clarify how the proposed methodological approach is appropriate for the claims made? 
    \answerYes{While we do not make any claims outside of describing the dataset, we do clearly outline the methodology for collecting and describing the data in the sections entitled ``Collection Methods'' and ``Data Description''.}
   \item Do you clarify what are possible artifacts in the data used, given population-specific distributions?
    \answerNA{NA}
  \item Did you describe the limitations of your work?
    \answerNA{NA}
  \item Did you discuss any potential negative societal impacts of your work?
    \answerYes{Yes. We discuss potential negative uses of this data in the section entitled ``Data Use Guidance''.}
      \item Did you discuss any potential misuse of your work?
    \answerYes{Yes. We provide guidance for responsible use of the data in the section entitled ``Data Use Guidance''.}
    \item Did you describe steps taken to prevent or mitigate potential negative outcomes of the research, such as data and model documentation, data anonymization, responsible release, access control, and the reproducibility of findings?
    \answerYes{Yes. We have provided responsible guidelines for the data's use and have anonymized the data as much as possible without detrimenting provenance and usefulness. For example, we anonymized the IP address of these outlets.}
  \item Have you read the ethics review guidelines and ensured that your paper conforms to them?
    \answerYes{Yes, we have carefully reviewed the ethics guidelines and ensured our paper and the release of this dataset conforms to them.}
\end{enumerate}

\item Additionally, if your study involves hypotheses testing...
\begin{enumerate}
  \item Did you clearly state the assumptions underlying all theoretical results?
    \answerNA{NA}
  \item Have you provided justifications for all theoretical results?
    \answerNA{NA}
  \item Did you discuss competing hypotheses or theories that might challenge or complement your theoretical results?
    \answerNA{NA}
  \item Have you considered alternative mechanisms or explanations that might account for the same outcomes observed in your study?
    \answerNA{NA}
  \item Did you address potential biases or limitations in your theoretical framework?
    \answerNA{NA}
  \item Have you related your theoretical results to the existing literature in social science?
    \answerNA{NA}
  \item Did you discuss the implications of your theoretical results for policy, practice, or further research in the social science domain?
    \answerNA{NA}
\end{enumerate}

\item Additionally, if you are including theoretical proofs...
\begin{enumerate}
  \item Did you state the full set of assumptions of all theoretical results?
    \answerNA{NA}
	\item Did you include complete proofs of all theoretical results?
    \answerNA{NA}
\end{enumerate}

\item Additionally, if you ran machine learning experiments...
\begin{enumerate}
  \item Did you include the code, data, and instructions needed to reproduce the main experimental results (either in the supplemental material or as a URL)?
    \answerNA{NA}
  \item Did you specify all the training details (e.g., data splits, hyperparameters, how they were chosen)?
    \answerNA{NA}
     \item Did you report error bars (e.g., with respect to the random seed after running experiments multiple times)?
    \answerNA{NA}
	\item Did you include the total amount of compute and the type of resources used (e.g., type of GPUs, internal cluster, or cloud provider)?
    \answerNA{NA}
     \item Do you justify how the proposed evaluation is sufficient and appropriate to the claims made? 
    \answerNA{NA}
     \item Do you discuss what is ``the cost`` of misclassification and fault (in)tolerance?
    \answerNA{NA}
  
\end{enumerate}

\item Additionally, if you are using existing assets (e.g., code, data, models) or curating/releasing new assets, \textbf{without compromising anonymity}...
\begin{enumerate}
  \item If your work uses existing assets, did you cite the creators?
    \answerYes{Yes. NELA-PS was created using an existing list of pink slime outlets created by \cite{bengani2019hundreds, bengani2020election}. We have both cited the author and have made it explicit what portions of the dataset stem from the authors work.}
  \item Did you mention the license of the assets?
  \answerYes{Yes. We present the licensing of the released data in ``Distribution and Maintenance''. There is no license information for any existing assets used.}
  \item Did you include any new assets in the supplemental material or as a URL?
    \answerNA{NA}
  \item Did you discuss whether and how consent was obtained from people whose data you're using/curating?
    \answerNA{NA}
  \item Did you discuss whether the data you are using/curating contains personally identifiable information or offensive content?
    \answerYes{Yes. We have discussed the potential for offensive content in this data set and provide guidance related to that potential in the section entitled ''Data Use Guidance''.}
\item If you are curating or releasing new datasets, did you discuss how you intend to make your datasets FAIR?
\answerYes{Yes. Under the section entitled ``FAIR Principles'', we discuss how we made our dataset FAIR.}
\item If you are curating or releasing new datasets, did you create a Datasheet for the Dataset? 
\answerNo{No. While we did not create a separate datasheet for the dataset, we made sure to cover the questions asked in \citet{gebru2021datasheets}. In particular, we cover these questions in the section entitled ``Data Formats and Distribution''.}
\end{enumerate}

\item Additionally, if you used crowdsourcing or conducted research with human subjects, \textbf{without compromising anonymity}...
\begin{enumerate}
  \item Did you include the full text of instructions given to participants and screenshots?
    \answerNA{NA}
  \item Did you describe any potential participant risks, with mentions of Institutional Review Board (IRB) approvals?
    \answerNA{NA}
  \item Did you include the estimated hourly wage paid to participants and the total amount spent on participant compensation?
    \answerNA{NA}
   \item Did you discuss how data is stored, shared, and deidentified?
   \answerNA{NA}
\end{enumerate}


\end{enumerate}

\end{document}